\documentclass[fleqn,10pt]{wlscirep}
 \usepackage{dcolumn}% Align table columns on decimal point
 \usepackage{bm}% bold math
 \usepackage{natbib}

 \def\be{\begin{equation}}
 \def\ee{\end{equation}}
 \def\bea{\begin{eqnarray}}
 \def\eea{\end{eqnarray}}
 \def\kv{{\bf k}}

 \def\no{\nonumber}
 \def\ra{\rangle}
 \def\la{\langle}
 \def\s{\sigma}
 \def\t{\tau}

 \usepackage{color}

 \newcommand{\dg}[1]{{#1}^\dagger}

\title{Spin supersolid phase in coupled alternating spin chains}

\author[1,2]{F. Heydarinasab}
\author[1,+]{J. Abouie}
\affil[1]{Department of Physics, Institute for Advanced
    Studies in Basic Sciences (IASBS), Zanjan 45137-66731, Iran}
\affil[2]{Department of Physics,
    Faculty of Science, University of Sistan and Baluchestan, Zahedan,
    Iran}

  \affil[+]{jahan@iasbs.ac.ir}

 \begin{abstract}
We study the ground state phase diagram of a two dimensional
mixed-spin system of coupled alternating spin-1 and $1/2$ chains
with a stripe supersolid phase. Utilizing different analytical and
numerical approaches such as mean field approximation, cluster mean
field theory and linear spin wave theory, we demonstrate that our
system displays a rich ground state phase diagram including novel
stripe supersolid, solids with different fillings and
super-counterfluid phases, in addition to a stripe solid with half
filling, superfluid and Mott insulating phases. In order to find a
minimal mixed-spin model for stripe supersolidity, in the second
part of the paper we consider two kinds of mixed-spin system of
coupled alternating spin-1 and $1/2$ chains with (i) anisotropic
nearest neighbor interactions, (ii) anisotropic hoppings and study
their ground state phase diagrams. We demonstrate that, for the
systems with uniform hoppings, the repulsive intra-chains
interactions are necessary for stripe supersolidity. In this case
the minimal two dimensional mixed-spin model is a system of spin-1
and spin-1/2 XXZ chains, interacting via a XY Hamiltonian. In the
case of anisotropic hoppings, a system of coupled Ising chains is
the minimal model.
\end{abstract}

\begin{document}

    \flushbottom
    \maketitle

\thispagestyle{empty}

    \section*{Introduction}\label{sec:intro}
    Supersolids are characterized by the coexistence of diagonal solid
    and off-diagonal superfluid long-range orders\cite{thouless1969flow,
        andreev1971quantum,Matsuda701, Liu1973}. Combination of these two
    apparently antithetical properties has attracted the attentions of
    both experimentalists and theorists, and searching for this exotic
    phenomenon has become one of the main subjects of condensed matter
    and cold atoms physics\cite{kim2004probable, kim2004observation,
        li2017stripe, leonard2017supersolid}. Since the pioneering work of
    Jaksch {\it et al.}, in describing the dynamics of an ultracold
    dilute gas of bosonic atoms in optical lattices with a Bose-Hubbard
    model\cite{jaksch1998cold}, lots of efforts have been devoted to
    search for supersolid phases in Bose-Hubbard models, experimentally
    and theoretically on one dimensional (1D)
    chains\cite{PhysRevLett.97.087209, PhysRevA.79.011602, PhysRevB.80.174519, PhysRevLett.88.170406}, two dimensional (2D)
    \cite{PhysRevLett.88.170406, 0295-5075-72-2-162,
        PhysRevLett.94.207202, PhysRevLett.95.033003, PhysRevLett.98.260405,
        PhysRevLett.103.225301, PhysRevLett.104.125301, PhysRevA.82.013645,
        Mila01062008, PhysRevB.86.054516, PhysRevB.75.174301,
        PhysRevB.75.214509, PhysRevLett.95.127205, PhysRevLett.95.127206,
        PhysRevLett.95.127207, PhysRevLett.95.237204, PhysRevB.76.144420,
        PhysRevLett.100.147204, PhysRevLett.104.125302, PhysRevB.84.054510,
        PhysRevB.84.174515, PhysRevA.85.021601, PhysRevLett.97.147202}
    lattice structures, bilayer systems of dipolar lattice
    bosons\cite{PhysRevLett.103.035304, trefzger2010quantum} and three
    dimensional (3D) cubic  lattices\cite{PhysRevLett.98.260405,
        PhysRevB.79.094503, PhysRevB.84.054512, PhysRevLett.108.185302}.

    Another appropriate ground for searching various supersolid phases
    are quantum spin systems. It has been shown that 1D spin-1
    chains\cite{sengupta2007spin, peters2009spin,rossini2011spin}, 2D
    frustrated spin-1/2\cite{PhysRevLett.97.127204, chen2010field,
        laflorencie2007quantum, PhysRevLett.100.090401,guo2014stripe,
        thomson2015bose, murakami2013supersolid, albuquerque2011phase,
        wierschem2013columnar, picon2008mechanisms, ng2017field,
        momoi2000magnetization} and spin-1\cite{sengupta2008ground,
        toth2012competition, su2014magnetic, sengupta2007field} models in an
    external magnetic field and 3D spin models\cite{ueda2013nematic,
    selke2013multicritical} possess different kinds of stripe,
    checkerboard and star supersolid phases. However, in spite of many
    studies on uniform spin systems, the supersolidity of mixed-spin
    systems has not been addressed so far. Mixed-spin systems, or
    quantum ferrimagnets which are composed of different spins, mostly
    of two kinds, are a special class of spin models where their
    universality class is completely different from uniform spin
    models\cite{trumper2001antiferromagnetically, PhysRevB.70.184416, PhysRevB.73.014411, 1742-5468-2011-08-P08001}. Ferrimagnets, which
    occur rather frequently in nature, are somehow between the
    antiferromagnets and the ferromagnets. Their lowest energy band is
    gapless which shows a ferromagnetic behavior while there is a finite
    gap to the next band above it which has the antiferromagnetic
    properties. It is the acoustical and optical nature of excitations
    which is the result of two different types of spin in each unit
    cell.

    Recently, we have studied a 2D frustrated ferrimagnetic spin model,
    originating from an inhomogeneous 2D bosonic system, composed of two
    kinds of hard-core and semi-hard-core bosons with different
    nilpotency conditions, and shown that the model on a square lattice
    with nearest-neighbor (NN) and next-nearest-neighbor (NNN)
    interactions displays the checkerboard supersolid
    phase\cite{heydarinasab2017inhomogeneous} which is not observed in the 2D uniform
    spin-$\frac 12$ system on square lattices with short-range
    interactions\cite{PhysRevB.86.054516, hebert2001quantum,
        PhysRevLett.84.1599}. Actually the interactions between spins with
    different sizes decrease the quantum fluctuations and cause the
    stabilization of the checkerboard supersolid order. In this paper,
    we introduce a different system of coupled alternating spin $\t=1$
    and $\s=\frac 12$ chains (CAS) (see Fig. \ref{fig:ferri} and Eq.
    (\ref{Hamiltonian})) and show that our CAS system possesses a stripe
    supersolid (STS) phase, characterizing by the coexistence of stripe
    solid (ST) and superfluid (SF) orders. We investigate the ground
    state phase diagram of the CAS model using different analytical and
    numerical approaches such as mean field (MF) approximation, cluster
    mean field theory (CMFT) and linear spin wave theory (LSWT).
    Competition between NN and NNN interactions causes the system to
    undergo various first- and second-order phase transitions, and
    different solids, Mott insulators (MI), SF and super-counterfluid
    (SCF) to appear in the ground state phase diagram of the model, in
    addition to the STS. By studying the behavior of spin wave
    excitations, we investigate the stability of MF orders and
    demonstrate that, except at the superfluid-supersolid transition
    lines, overall quantum fluctuations are small in our CAS system and
    the MF predictions concerning the stability of phases are reliable.

    In the second part of this paper, we look for a {\it minimal}
    mixed-spin CAS system, possessing an stable supersolid phase. In
    this respect, we consider two kinds of anisotropic CAS model: ($i$)
    a CAS system with anisotropic NN interactions where the intra-chains
    and inter-chains NN interactions are not the same, and ($ii$) a CAS
    system with anisotropic hopping energies in which the intra-chains
    and inter-chains hoppings are different. By obtaining the CMFT
    ground state phase diagrams of these systems, we demonstrate that
    the appearance of the STS order strongly depends on the amounts of
    intra-chains NN interaction.  By studying the behavior of spin wave
    excitations, and also the behavior of diagonal and off-diagonal
    order parameters by CMFT with larger cluster sizes, we investigate
    amount of quantum fluctuations and consequently the stability of the
    STS phase in all anisotropic CAS systems. Our results indicate that
    the repulsive intra-chains NN interactions are necessary for the
    emergence of the STS phase. Finally, based on our achievements, we
    present a minimal mixed-spin CAS model with stable supersolid phase
    in the ground state phase diagram.

    Our CAS model could be related to the mixed-valance iron
    polymer\cite{zheng2009spin} in the spin system and ladder-like
    optical lattices \cite{li2013topological} in the bosonic systems.
    Also our model could be realized in coupled one dimensional optical
    lattices\cite{safavi2014quantum} by alternatively changing the
    optical depth.

    This paper is organized as follows. In Sec. \ref{sec:model} we
    introduce our CAS system and present the bosonic counterpart of this
    model. In Sec. \ref{sec:GPD} we obtain the ground state phase
    diagram of the CAS model using MF approximation, CMFT and LSWT. The
    linear spin wave dispersions and number of excitation modes, are
    also presented in this section. The anisotropic CAS models with
    anisotropic NN interactions, and anisotropic hoppings are
    investigated in Sec. \ref{sec:ani-CAS}. Finally, we will summarize
    our results and give the concluding remarks in Sec.
    \ref{sec:summary}.

%%%%%%%%%%%%%%%%%%%%%%%%%%%%%%%%%%%%%%%%%%%
\section{Our Model}\label{sec:model}
\subsection {Coupled alternating spin 1 and $\frac 12$ chains}
\begin{figure}[h]
    \centerline{\includegraphics[width=70mm]{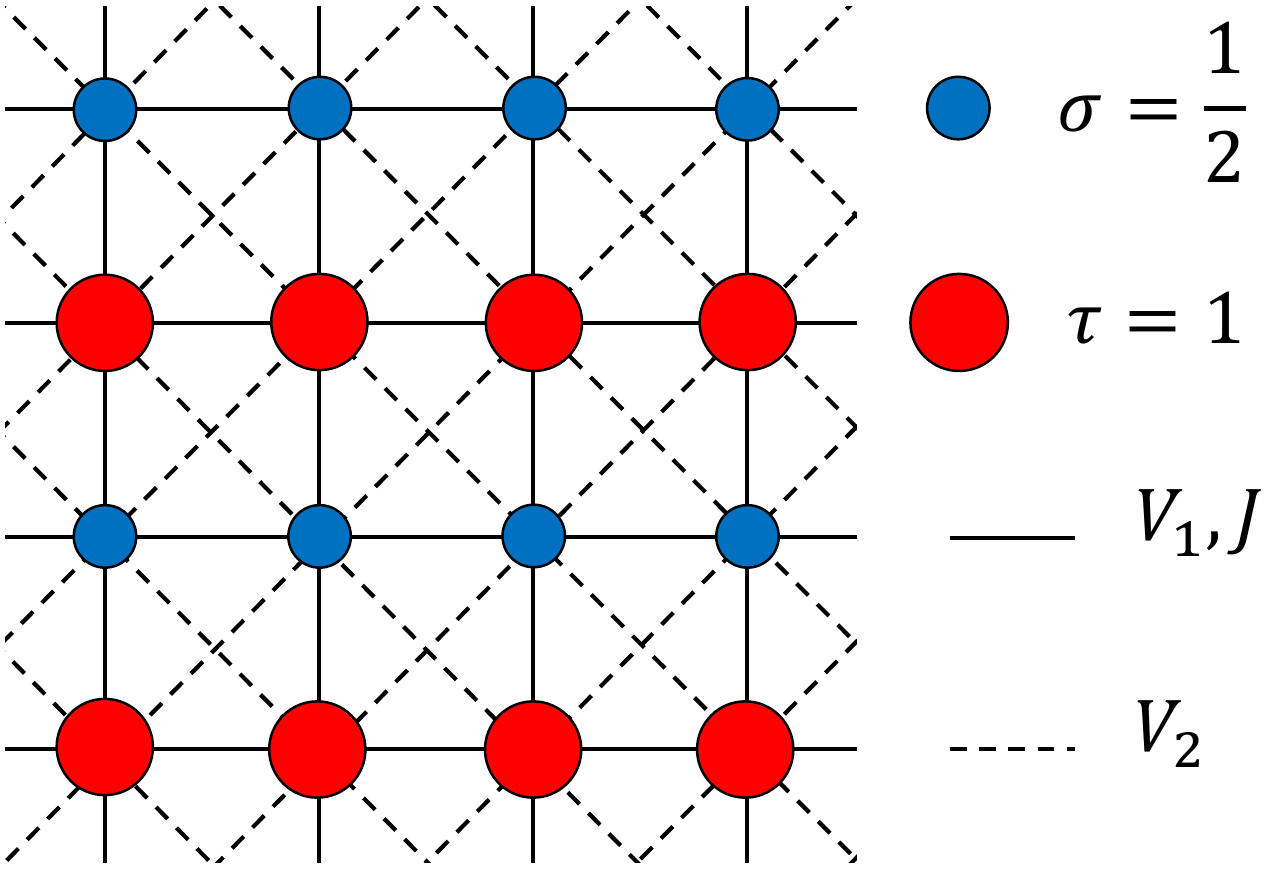}} \caption{(Color
        online) The schematic illustration of the coupled alternating spin-1
        and $1/2$ chains. The small (large) circles show spin $\s=\frac 12$
        ($\t=1$). The NN and NNN interactions are depicted by the solid and
        dashed lines, respectively.} \label{fig:ferri}
\end{figure}

Let us consider a 2D system of coupled alternating spin $\t=1$ and
$\s=\frac 12$ chains (CAS), describing by the following Hamiltonian:
\begin{equation}
H=H_{\s}+H_{\t}+H_{\s\t}, \label{Hamiltonian}
\end{equation}
with
\begin{eqnarray}
\nonumber H_\s&=&\sum_{\la i,j\ra, \alpha}{\cal J}^\alpha\s_i^\alpha \s_j^\alpha -h\sum_i \s_i^z,\\
\nonumber H_\t&=&\sum_{\la i,j\ra, \alpha}{\cal J}^\alpha\t_i^\alpha \t_j^\alpha -h\sum_i \t_i^z,\\
\nonumber H_{\s\t}&=&\sum_{\la i,j\ra, \alpha}{\cal
    J}^\alpha\s_i^\alpha \t_j^\alpha +V_2\sum_{\la\la
    i,j\ra\ra}\s_i^z\t_j^z,
\end{eqnarray}
where $\alpha=x, y$ and  $z$. The Hamiltonians $H_\s$, $H_\t$, and
$H_{\s\t}$ include the intra-chains and inter-chains interactions,
respectively. The parameters ${\cal J}^{x,y}(=-2J)$ and ${\cal
    J}^z(=V_1)$ are the NN interactions, $V_2$ denotes the inter-chains
NNN interactions, and $h$ is a magnetic field along $z$ direction.
The magnetic field $h$ is proportional to applied magnetic fields as
$h=g_\s\mu_{\rm B} B_\s=g_\t\mu_{\rm B} B_\t$, where $\mu_{\rm B}$
is the Bohr magneton, $g_\s$ and $g_\t$ are the g-factors,
respectively, for spins-$1/2$ and spins-1, and $B_\s$ and $B_\t$ are
the external magnetic fields applied to the subsystems with spins
$1/2$ and $1$, respectively. Throughout this paper we consider $g_\s
B_\s=g_\t B_\t$, and study the effects of a uniform $h$ on the
ground state phase diagram of the system. Our CAS system is
schematically shown in Fig. \ref{fig:ferri}.
%%%%%%%%%%%%%%%%%%%%%%%%%% FIG 0 %%%%%%%%%%%%%%%%%%
%%%%%%%%%%%%%%%%%%%%%%%%%%%%%%%%%%%%%%%%%%%%%%%%%

The Hamiltonian in Eq. (\ref{Hamiltonian}) possesses the
translational symmetry of the 2D lattice with the translational
vector $a\hat{x}+2a\hat{y}$ as well as the rotational U(1) symmetry.
Spontaneously breaking of these symmetries by varying the model
parameters, causes the system to experience various first- and
second-order phase transitions. Different diagonal and off-diagonal
long-range orders appear in the ground state phase diagram of the
above model which will be discussed in next sections.

%%%%%%%%%%%%%%%%%%%%%% Bosonic Counterpart %%%%%%%%%%%%%%%%%%%%
\subsection{Bosonic counterpart of the CAS chains}\label{sec:Bosonic}

In bosonic language, the systems of coupled uniform spin-1/2 chains
are equivalent to the systems of coupled 1D optical lattices,
containing hard-core bosons. These systems possess different
superfluid and solid phases\cite{safavi2014quantum}, in the presence
of intra-chains hoppings, repulsive intra-chains and attractive
inter-chains interactions, but no supersolid phase is formed due to
the hard-core nature of the bosons.

The three-body-constrained bosons (or semi-hard-core bosons) with
the nilpotency condition $(b_i^\dagger)^3=0$, may remove this
problem. This nilpotency condition signifies that one can put up two
$b$ particles on each lattice site. Recent studies on the coupled 1D
optical lattices containing two kinds of boson, a hard-core and a
semi-hard-core boson, show that the system displays different MIs
and SF orders \cite{singh2017quantum}, in the presence of
intra-chains hoppings and inter-chains interactions, but still solid
and supersolid phases are absent in the ground state phase diagram.
Indeed in the absence of intra-chains repulsive interactions the
translational symmetry of the system preserves and consequently no
solidity occurs in the system. So, considering repulsive
intra-chains interactions on the semi-hard-core bosons can result in
a supersolid order in the system of coupled 1D optical lattices.

Here, by using the relations between semi-hard-core bosons and
spin-1 operators\cite{heydarinasab2017inhomogeneous}, we map our CAS model in Eq.
(\ref{Hamiltonian}) to a bosonic system of coupled alternating
hard-core and semi-hard-core bosonic lattices. We demonstrate that
the presence of repulsive intra-chains interactions is sufficient
for the appearance of solid and supersolid phases.

Let us consider a system of coupled alternating 1D optical lattices
with hard-core and semi-hard-core bosons, $a$ and $b$, which
interact via the following Hamiltonian:
\begin{eqnarray}
H&=& H_a+H_b+H_{ab},\\
\no H_a&=& \sum_{\la i,j\ra}[(t^{a} \dg a_i a_j+h.c.)+V^{a} n^a_i n^a_j]- \sum_i \mu^a n^a_i,\\
\no H_b&=& \sum_{\la i,j\ra}[(t^{b} \dg b_i b_j+h.c.)+V^{b} n^b_i n^b_j]- \sum_i  \mu^b n^b_i,\\
\no H_{ab}&=& \sum_{\la i,j\ra}[(t^{ab} \dg a_i b_j+h.c.)+V^{ab}
n^a_i n^b_j]+V_2\sum_{\la\la i,j\ra\ra} n^a_i n^b_j,\label{BH1}
\end{eqnarray}
where $H_a$ ($H_b$) contains interactions between $a$ ($b$) bosons
and $t^{a}$ ($t^{b}$), $V^{a}$ ($V^{b}$) and $\mu^a$ ($\mu^b$) are
the hopping energy, the interaction and the chemical potential in
$a$ ($b$) sublattice. $H_{ab}$ gives the interactions between $a$
and $b$ bosons with $t^{ab}$ the hopping energy and $V^{ab}$, and
$V_2$ the interaction energies. $\dg a_i$ $(a_i)$ and $\dg b_j$
$(b_j)$ are respectively the creation (annihilation) operators of
$a$ and $b$ particles at sites $i$ and $j$, on a 2D square lattice.
The particles $a$ are canonical hard-core bosons and satisfy the
canonical commutation relations. The number of $a$ bosons at site
$i$ is $n^a_i=\dg a_i a_i$, and the nilpotency condition for these
bosons is $(a_i^\dagger)^2=0$. The $b$ particles are semi-hard-core
bosons and satisfy the following statistics algebra:
\begin{eqnarray}
\no &&[b_i,b_j]=[\dg b_i, \dg b_j]=0,\\
&&[b_i, \dg b_j]=\delta_{ij}(1-n^b_i),~~~~ [n^b_i, \dg
b_j]=\delta_{ij}\dg b_j, \label{commutation-b}
\end{eqnarray}
where $n^b_i$ is the number of $b$ bosons which possesses the
relation $\dg {(n^b_i)}=n^b_i$.

In order to obtain the bosonic Hamiltonian in Eq. (\ref{BH1}) we
have used the following linear spin-boson transformations between
$a$ bosons and spin-$\frac 12$\cite{Matsubara1956}, and between $b$
bosons and the spin-1 operators\cite{Batista2004}:
\begin{eqnarray}
\no && \s_i^z=n_i^a-\frac 12,~~~\s_i^+=\dg a_i,~~~\s_i^-=a_i ,\\
&&\tau_j^z=n^b_j-1,~~~ \tau_j^+=\sqrt{2}\dg b_j,~~~
\tau_j^-=\sqrt{2} b_j.\label{eq:mappings}
\end{eqnarray}

Since these spin-boson transformations are isomorphic, all
symmetries and physical properties of the CAS system
(\ref{Hamiltonian}) and the bosonic system (\ref{BH1}) are
identically the same. The bosonic Hamiltonian in Eq. (\ref{BH1}) is
transformed to the mixed-spin Hamiltonian in Eq. (\ref{Hamiltonian})
by defining the following relations:
\begin{eqnarray}
\no && V^{a(b)}=V^{ab}=V_1,\\
&& t^{a} \to J,~t^{b} \to J/2,~ t^{ab} \to J/\sqrt 2,\\
\no && \mu^a \to h+4V_1+4V_2,~ \mu^b \to h+5V_1+2V_2.
\end{eqnarray}

%%%%%%%%%%%%%%%%%%%%%%%% ground state phase diagram %%%%%%%%%%%%%%%%%%%%%%%%
\section{Ground state phase diagram}\label{sec:GPD}

In order to obtain the ground state phase diagram of the CAS model,
first we use a MF approximation to investigate the system
classically and then utilizing a generalized CMFT we obtain the
modified ground state phase diagram of the CAS model.
%%%%%%%%%%%%%%%%%%%%%% Definition of the orders %%%%%%%%%%%%%%%%%%%%
\begin{table}[ht]
    \centering
    \caption{Definitions of various orders. $M_v$ is the total transverse
        magnetization.
        The fillings (average number of bosons in each unit cell) are mentioned in the parentheses.}
    \label{tab:Lorders}
    \begin{tabular}{|l|l|l|}
        \hline
        Phases&sublattices magnetizations&$M_v$\\
        \hline
        ST(3/6)& $m_A^z=-m_B^z$, ~$M_C^z=-M_D^z$& 0\\
        \hline
        $b$ST(4/6)& $m_A^z=m_B^z$, ~$M_C^z=-M_D^z$& 0\\
        \hline
        MI(4/6)& $m_A^z=m_B^z$, ~$M_C^z=M_D^z$& 0\\
        \hline
        $a$ST(5/6)& $m_A^z=-m_B^z$, ~$M_C^z=M_D^z$& 0\\
        \hline
        Full& $m_A^z=m_B^z=1/2$, ~$M_C^z=M_D^z=1$& 0\\
        \hline
        STS& $m_A^z \neq m_B^z \neq M_C^z \neq M_D^z$& $\neq 0$ \\
        \hline
        SF& $m_A^z=m_B^z$, ~$M_C^z=M_D^z$& $\neq 0$\\
        \hline
        SCF& $m_A^z=m_B^z$, ~$m_A^{x,y}=-m_B^{x,y}\neq 0$, $M_C^z=M_D^z$,  ~$M_C^{x,y}=-M_D^{x,y}\neq 0$& 0\\
        \hline
    \end{tabular}
\end{table}
%%%%%%%%%%%%%%%%%%%%%%%%%% FIG 2 %%%%%%%%%%%%%%%%%%
\begin{figure}[ht!]
    \centering
    \includegraphics[width=120mm]{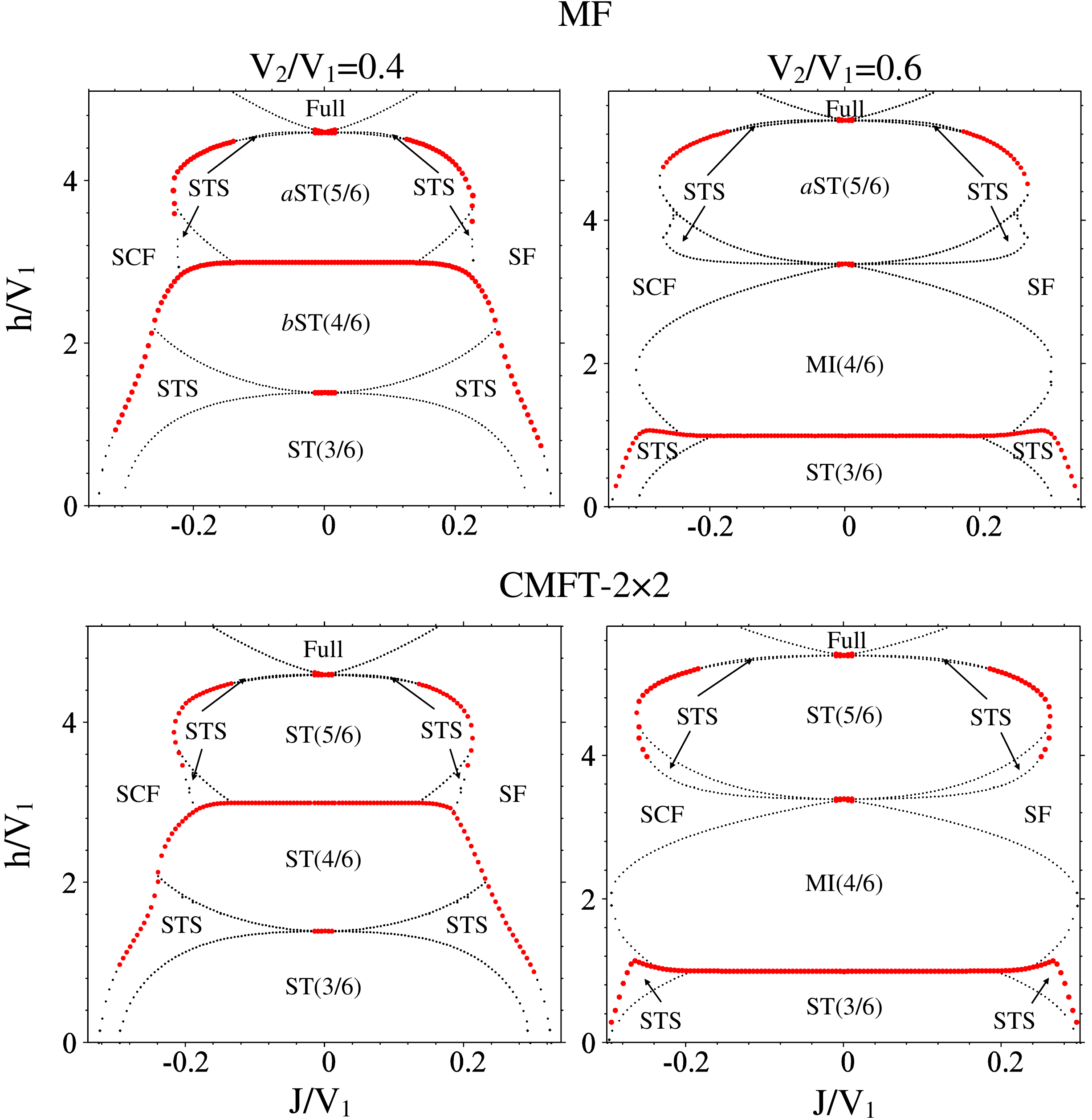}\\
    \includegraphics[width=110mm]{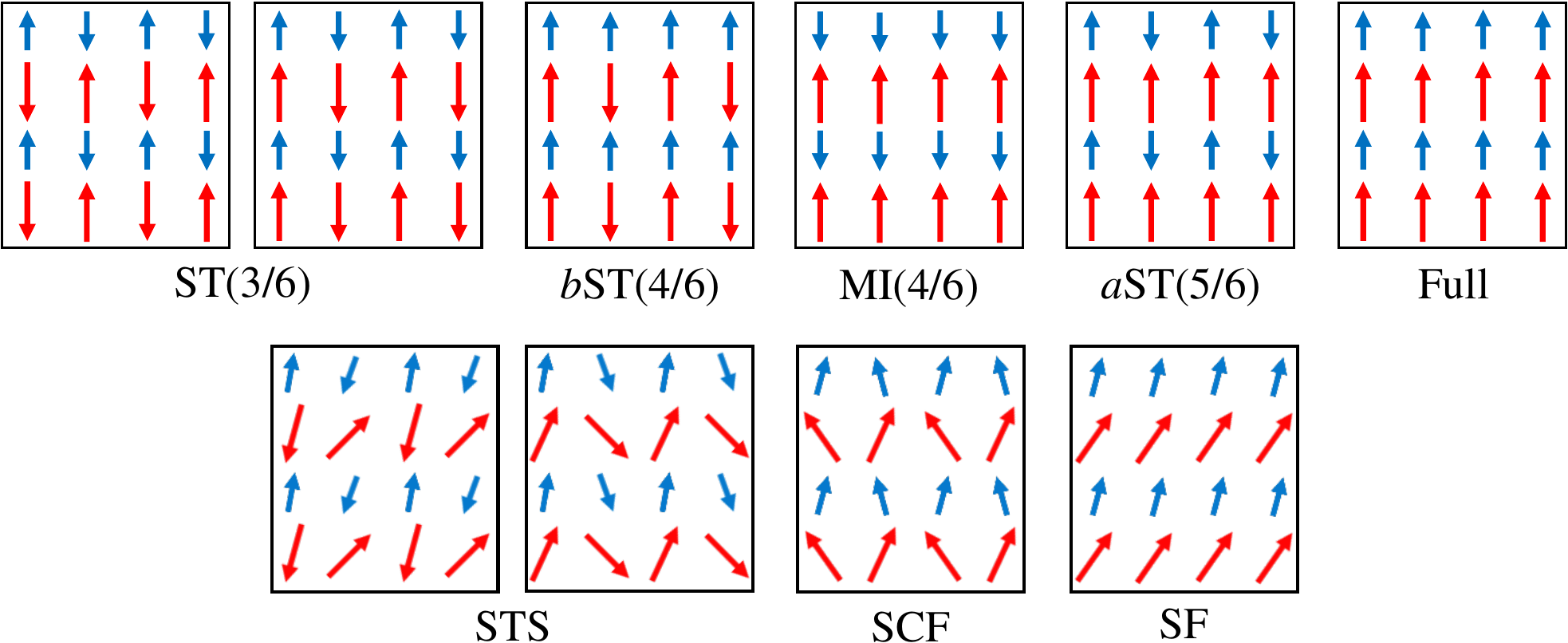}
    \caption{(Color online) Ground state phase diagrams of the CAS system for the two different strengths of frustration: $\frac{V_2}{V_1}=0.4$ and
        $\frac{V_2}{V_1}=0.6$. Top and middle: MF and CMFT phase diagrams.
        The red (black) dotted lines show first-order (second-order) phase transitions.
        Bottom: schematic illustrations of various solids, SF, SCF and STS phases.
        The spins alignment in the left (right) panel of ST(3/6) phase
        occurs for $V_2/V_1<0.5$ ($>0.5$) to satisfy the interaction $V_1$
        ($V_2$). Moreover, the spins alignment in the left (right) panel of
        the STS phase forms around the left (right) side of the ST(3/6)
        phase. }
    \label{fig:MFphase-diag-U0}
\end{figure}
%%%%%%%%%%%%%%%%%%%%%%%%%%%%%%%%%%%%%%%%%%%%%%%%%%%%%%%%%%%%%%

In the MF approximation, we first divide the system into four
sublattices, the subsystem with spins $\s$ into A and B, and the
subsystem with spins $\t$ into C and D, and then approximate the
local spins averages with MF order parameters. The four-sublattice
structure is expected to be emerged due to the NN and NNN
interactions. By defining the MF order parameters:
$\la\s_{i_A}^{\alpha}\ra=m_A^\alpha$,
$\la\s_{i_B}^{\alpha}\ra=m_B^\alpha$,
$\la\t_{i_C}^{\alpha}\ra=M_C^\alpha$, and
$\la\t_{i_D}^{\alpha}\ra=M_D^\alpha$, where $\la\dots\ra$ denotes
the expectation value on the MF ground state, the Hamiltonian in Eq.
(\ref{Hamiltonian}) is simply transformed to a single particle MF
Hamiltonian. The expectation values of spin operators on the ground
state of the MF Hamiltonian (sublattices' magnetizations) are given
in terms of other MF order parameters. Various kinds of long-range
diagonal and off-diagonal orders are defined by these
magnetizations. In the table \ref{tab:Lorders}, we have defined all
possible phases, appearing in the MF ground state phase diagrams.

In order to obtain the more accurate phase diagrams, we employ CMFT
with different cluster sizes as a more precise approach. CMFT is an
extension of the standard MF approximation in which clusters of
multiple sites are used as an approximate system instead of single
sites \cite{PhysRevB.86.054516, PhysRevA.85.021601,
    yamamoto2009correlated, heydarinasab2017inhomogeneous}. Recently we have generalized CMFT
for the staggered 2D mixed-spin system\cite{heydarinasab2017inhomogeneous}. Treating
exactly the interactions within the cluster and including the
interaction of spins outside the cluster as an effective field, one
can partially take into account fluctuations around classical ground
state as well as the effects of correlations of particles. Similar
to our MF analysis, we assume a background with four-sublattice
structure and embed a cluster of $N_C$ sites into this background.
Instead of treating the many-body problem in the whole system, we
consider the following effective cluster Hamiltonian:
\begin{equation}
H_C^{eff}=H_C+H_{\bar C}
\label{C-Hamiltonian},
\end{equation}
where the interactions within clusters are given by $H_C$ while the
interactions of spins inside the clusters with the rest of the
system are included in $H_{\bar C}$. The Hamiltonian $H_C$ is given
by Eq. (\ref{Hamiltonian}) where the summations run over sites $i,j
\in C$, and the Hamiltonian $H_{\bar C}$ is given in terms of CMFT
ground state magnetizations, which are computed self-consistently.

We have plotted in Figs. \ref{fig:MFphase-diag-U0}-top, and
\ref{fig:MFphase-diag-U0}-middle the MF and CMFT ground state phase
diagrams of the CAS system, for the two different strengths of
frustration: $V_2/V_1=0.4$ and $0.6$. We have also illustrated the
schematic pictures of various orders, at the bottom of Fig.
\ref{fig:MFphase-diag-U0}. For small values of $J/V_1$, independent
of the strengths of frustrations, the ground state phase diagram is
symmetric with respect to the $J=0$ line. Far from $J=0$ line, the
system, however, behaves differently for $J>0$ and $J<0$ regions.
For large values of $J>0$ the U(1) symmetry of the system breaks
spontaneously, and the SF long range order emerges in the system
where each boson is spread out over the entire lattice, with
long-range phase coherence. For $J<0$, instead of SF phase, the SCF
phase appears in the phase diagram, where the transverse components
of the spins lie in opposite directions (see Fig.
\ref{fig:MFphase-diag-U0}, the schematic picture of SCF). This phase
is characterized by a transverse staggered magnetization and a
longitudinal magnetization\cite{kuklov2003counterflow}. In bosonic
language the SCF order parameter is given by $\la a_i b_j^\dag \ra$
\cite{kaurov2005drag}. Although SCF is not a superfluid, but as we
will show by means of LSWT, its excitation spectrum is identically
the same as the SF phase. In this phase due to the transverse
staggered magnetization the easy plane U(1) symmetry reduces to the
$Z_2$ one\cite{chen2010quantum} and the translational symmetry is
also broken. But, it is not a kind of solid since this phase
possesses a gapless excitation and the longitudinal staggered
magnetization is zero. This phase is not seen in the ground state
phase diagram of the 2D mixed-spin system with staggered arrangement
of spin-1/2 and spin-1\cite{heydarinasab2017inhomogeneous}. In the staggered 2D mixed-spin
system the ground state phase diagram is completely symmetric with
respect to $J=0$.
%%%%%%%%%%%%%%%%%%%%%%%%%%%%%%%%%%%%%%%%%%%%%%%%%%%%%%%%%%%%%%
%%%%%%%%%%%%%%%%%%%%%%%%%% FIG 3 %%%%%%%%%%%%%%%%%%
\begin{figure}[t]
        \centering
    \includegraphics[width=110mm]{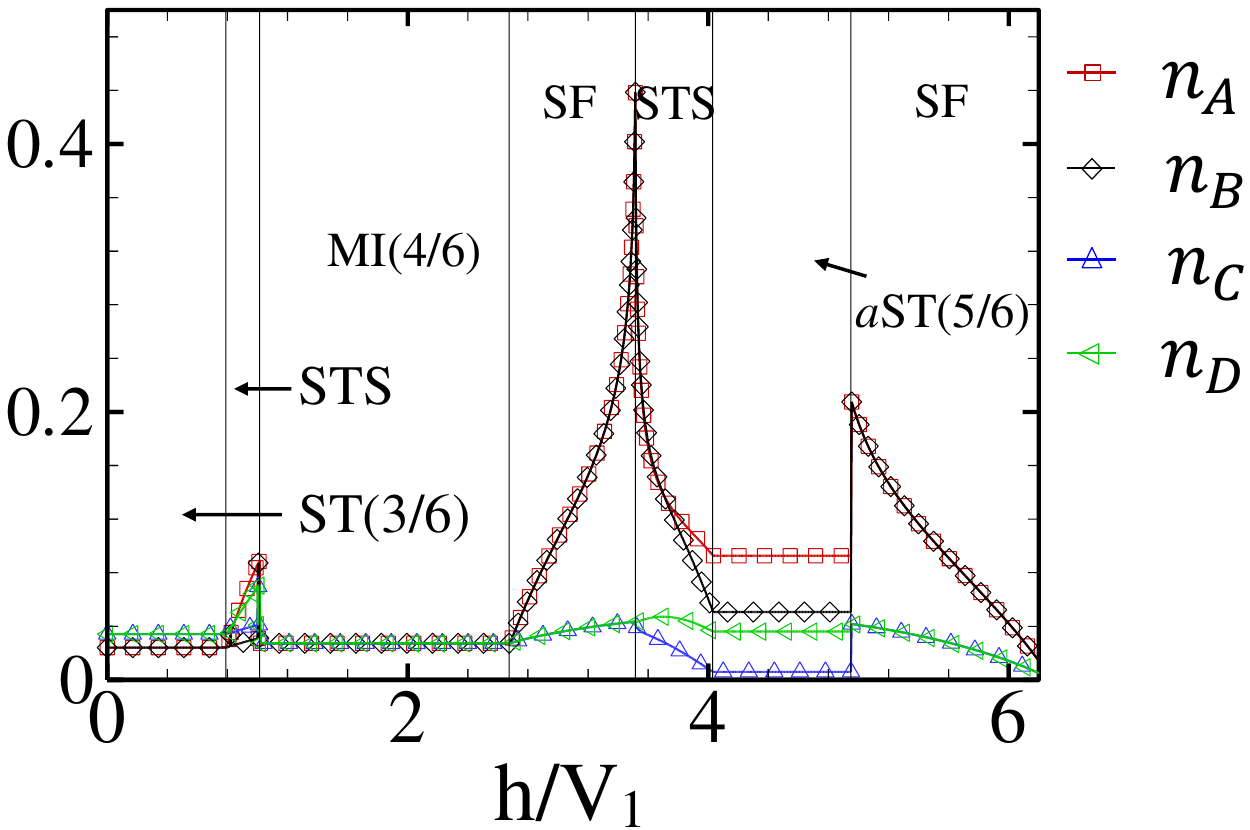}
    \caption{(Color online) Number of HP bosons on the MF ground state.
        $n_A$ and $n_B$ are the amount of quantum fluctuations in the subsystem with spin $\s$, and $n_C$ and $n_D$ are in the subsystem with spin $\t$.
        Overall quantum fluctuations are not strong enough to destroy MF orders, but for
        larger values of NNN interactions, the STS-SF transition lines are
        modified in comparison with the corresponding MF results.}
    \label{fig:SW-fluc}
\end{figure}
%%%%%%%%%%%%%%%%%%%%%%%%%%%%%%%%%%%%%%%%%%%%%%%%%%%%%%%%%%%%%

By decreasing $|J|$, at small magnetic field, aside from the U(1)
symmetry (which is completely broken in $J>0$ region and is
decreased to the $Z_2$ symmetry in $J<0$ one), the translational
symmetry of the system also breaks and a phase transition occurs
from SF and SCF to the STS phase in which both diagonal and
off-diagonal long range orders coexist in the system. The STS-SF and
STS-SCF phase transitions are of first- or second-order which are
attributed to the behavior of the low energy spin wave excitations.
Indeed, the abrupt and smooth changes of the low energy excitations
close to a transition point results in the discontinuous and
continuous variations of the diagonal and off-diagonal order
parameters.

In addition to the SF, SCF and STS phases, various kinds of stripe
solids: ST(3/6), $b$ST(4/6) and $a$ST(5/6), respectively with
fillings 3/6, 4/6, and 5/6 also appear in the phase diagram of the
CAS system. In these solid phases, depending on the values of
magnetic field, the translational symmetries of both subsystems or
one of them break spontaneously, (in $b$ST(4/6) the
translational symmetry of the subsystem with spin $\t$ and in
$a$ST(5/6) the translational symmetry of the subsystem with spin
$\s$). The stripe solid orders with fillings 4/6 and 5/6 are the
characteristics of our mixed-spin CAS system and are not seen in the
phase diagram of uniform XXZ spin-1/2 models. For $V_2/V_1=0.6$,
around $J=0$, instead of $b$ST(4/6), the system prefers to be in the
MI(4/6) phase at moderate magnetic field where both the
translational and U(1) symmetries are preserved in the system.
Actually, for larger values of $V_2/V_1$, the $V_2$ interactions try
to make the spins $\t$ and $\s$ antiparallel, such that the
translational symmetry of both subsystems preserves at moderate
magnetic field. This behavior that the translational symmetry does
not break even at large interactions, is the characteristic of the
two-component systems with inter-components interaction, which has
also been seen in the staggered mixed-spin system at
$V_2/V_1<0.5$\cite{heydarinasab2017inhomogeneous}.

Comparison between MF and CMFT phase diagrams shows that, for
$V_2/V_1<0.5$, there is no considerable changes in the MF phase
diagram in the presence of quantum fluctuations. The slight
deviations of the STS-SF transition lines, for $V_2/V_1>0.5$, at
large magnetic field, are attributed to the large amount of quantum
fluctuations at these boarders. In order to see the behavior of
quantum fluctuations in each phases, we utilize LSWT and study the
variations of spin waves' number in all sublattices. Using Holstein
Primakoff (HP) transformations, the Hamiltonian in Eq.
(\ref{Hamiltonian}) transforms to the following spin wave
Hamiltonian:
\begin{equation}
\label{SWH}\tilde{H}=E_0 + \sum_{\kv} \psi_\kv^\dag H_\kv \psi_\kv,
\end{equation}
where $E_0$ is the classical MF energy, $H_\kv$ is a square matrix in Fourier space, consisting the coefficients of bilinear terms, and $\psi_\kv$ is a
vector in terms of HP bosonic creation and annihilation operators.
Dimensions of $\psi_\kv$ and $H_\kv$ depend on the number of
sublattices in the MF ordered phases. Paraunitary diagonalization
\cite{colpa1978diagonalization} of $H_\kv$ yields the excitation
spectra in each phase, as well as the HP bosons' number.

Amount of quantum fluctuations in different phases is given by HP
bosons' numbers $n_A, n_B, n_C$ and $n_D$, respectively in the
sublattices $A$, $B$, $C$ and $D$.  As it is seen from Fig.
\ref{fig:SW-fluc}, in our mixed-spin model overall quantum
fluctuations are not strong enough to destroy the MF orders and the
MF predictions are reliable. However, at the second-order SF-STS and
first-order SF-$a$ST(5/6) transition lines, they are not negligible
and we should take them into account for reaching to the accurate
ground state phase diagrams. Also, since $n_A (n_C)$ is not equal to
$n_B (n_D)$, the $b$ST(4/6) and $a$ST(5/6) solids convert to the
ST(4/6) and ST(5/6) ones, respectively (see Fig.
\ref{fig:MFphase-diag-U0}, CMFT-($2\times 2$) phase diagram). It
should be noticed that although quantum fluctuations break the
translational symmetry of both subsystems but the fillings do not
change. Moreover, in the presence of quantum fluctuations, part of
the second-order STS-SF transition line below the ST(5/6) phase
transforms to a first order one. This means that the MF prediction
concerning the kind of transition order at this region is not
correct, and more precise approaches should be employed to obtain
the STS-SF critical and tricritical points.
%%%%%%%%%%%%%%%%%%%%%  2*4 results   %%%%%%%%%%%%%%%%%%%%%%%%%
\begin{figure}[t]
    \centering
    \includegraphics[width=60mm]{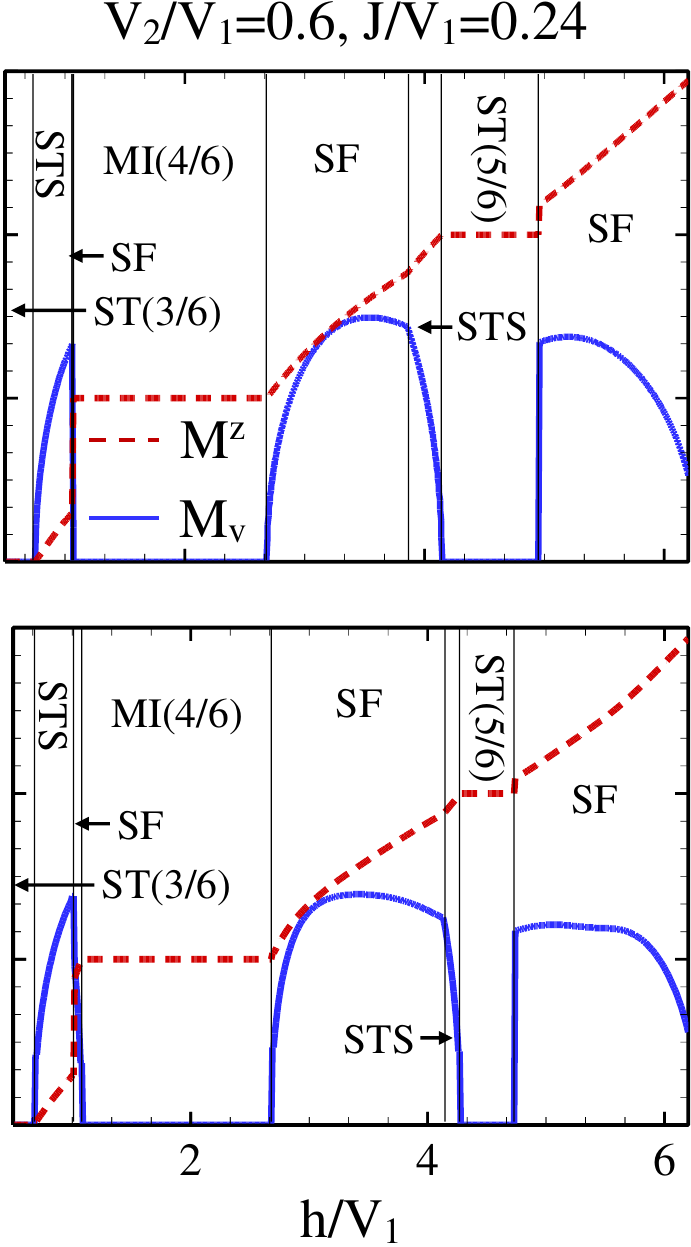}
    \caption{Diagonal and off-diagonal order parameters, computed by CMFT-$2\times 2$ (top) and CMFT-$2\times 4$ (bottom).}
    \label{fig:order-parameter}
\end{figure}
%%%%%%%%%%%%%%%%%%%%%%%%%%%%%%%  FigSW1   %%%%%%%%%%%%%%%%%%%%%%%%%%%
\begin{figure}[t]
    \centering
    \includegraphics[width=120mm]{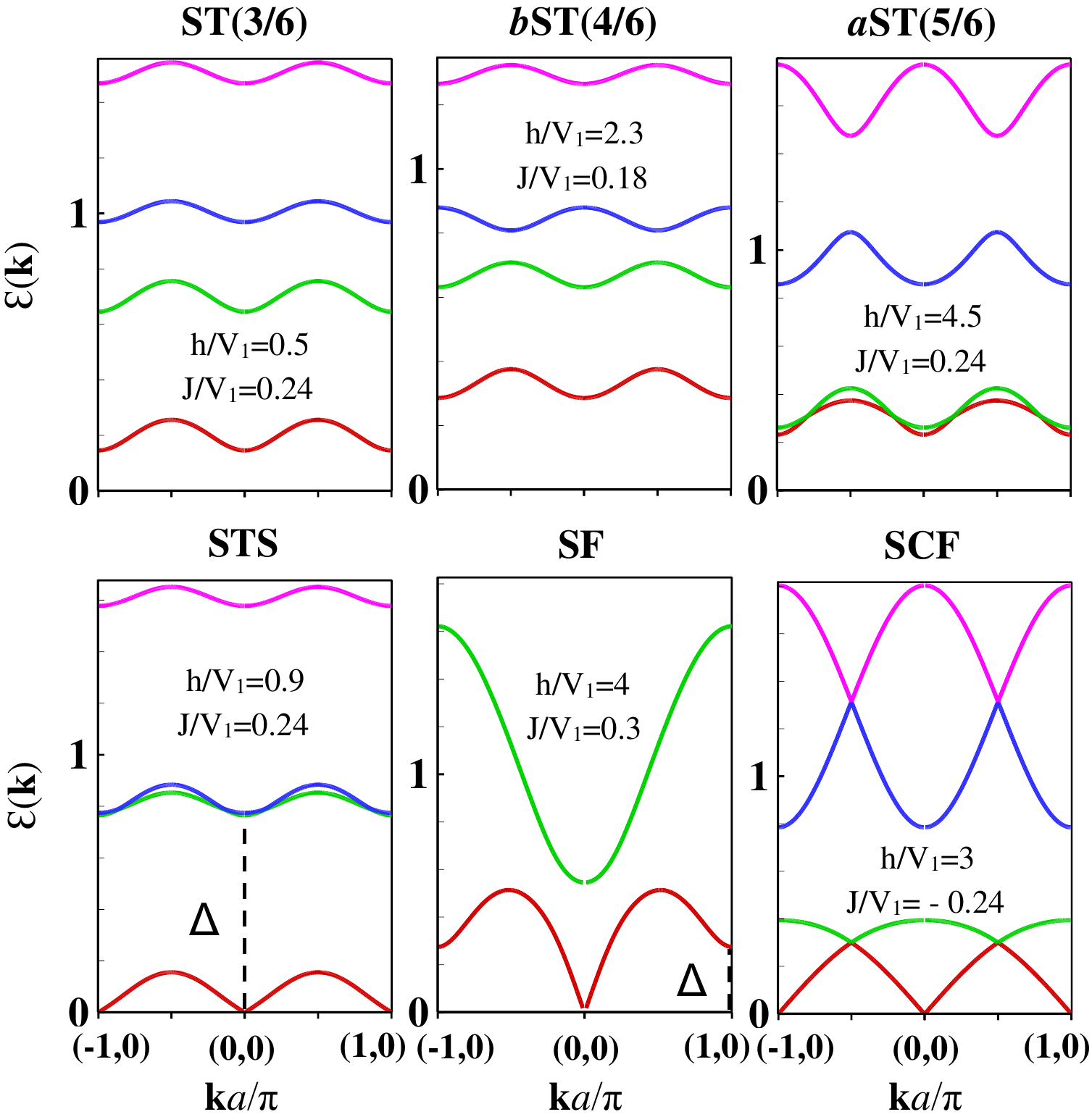}
    \caption{(Color online) Spin wave excitation spectra in various phases of the CAS system. Number of excitation modes reflects the number of sublattices in each phase.
        The parameter $\Delta$ in the SF and STS phases is the roton energy
        gap. In solid phases all excitations are gapped and the lowest spectrum
        has quadratic dispersion ($k^2$) around $\kv=(0,0)$. Whereas, in
        STS, SF and SCF phases, a gapless Goldstone mode appears in the
        excitation spectra.
        All plots are for $V_2/V_1=0.6$ except $b$ST(4/6) which is for $V_2/V_1=0.4$.} \label{fig:SW-Excitations}
\end{figure}
%%%%%%%%%%%%%%%%%%%%%%%%%%%%%%%%%%%%%%%%%%%%%%%%%%%%%%%%%%%%%%%%%%%%

We also investigate the behavior of both diagonal and off-diagonal
order parameters considering clusters with larger sizes in CMFT.
Employing clusters of eight spins [CMFT-($2\times 4$)], we have
computed the sublattices longitudinal and transverse magnetizations
for different values of $h$ and $J$ (see Fig.
\ref{fig:order-parameter}). Since quantum fluctuations are strong
around first order transition lines, some modifications around these
lines are expected. These are clearly seen by comparison of the
phases boarders of CMFT-($2\times 2$) and CMFT-($2\times 4$) in Fig.
\ref{fig:order-parameter}. The behavior of the order parameters
indicates that the STS phase around ST(5/6) solid becomes narrower
in the presence of quantum fluctuations. It seems that these regions
tend to be disappeared when we use CMFT with larger cluster sizes.
The STS phase appeared at small magnetic fields is however stable.
The instability of STS phase at larger magnetic field can be
explained as follows. The appearance of the STS phase is in fact the
result of the competition between the staggered magnetization along
$z$ direction (diagonal order) and the total transverse
magnetization (off-diagonal order). At large magnetic field, in the
presence of quantum fluctuations, both order parameters decrease but
the staggered magnetization is more sensitive and diminishes around
the phase boarder. So the STS region at larger magnetic field is
decreased in CMFT-($2\times 4$).

The CMFT becomes exact when the cluster size goes to infinity. In
practice, we are faced with computational limitations due to the
increasing of clusters' sizes and can not consider clusters of large
sizes. We thus should employ other techniques such as quantum Monte
Carlo simulations to obtain the exact phase diagram of the CAS
model.

We have also plotted in Fig. \ref{fig:SW-Excitations}, the spin wave
excitation spectra in all phases of the CAS system. Number of
excitation modes and their behavior depend on the number of
sublattices as well as their longitudinal and transverse
magnetizations. According to the translational symmetry of the CAS
system, the primitive vectors in the SF and MI phases are
$\vec{a}_1=a \hat{x}$ and $\vec{a}_2=2a \hat{y}$, and there are two
excitation modes in the system. However, when the translational
symmetry breaks in $x$ direction, as in the different solids, STS
and SCF phases, the primitive vectors are $\vec{a}_1=2 a\hat{x}$ and
$\vec{a}_2=2 a\hat{y}$ and the first BZ is folded in $x$ direction.
In these phases there exist four excitation modes in the system.

In STS and SF phases, as a result of the continues U(1) symmetry
breaking, a gapless Goldstone mode with a roton-like minimum
$\Delta$ appears in the excitation spectra (see Fig.
\ref{fig:SW-Excitations}). Appearance of the roton-like minimum in
the spectrum of these phases, is the characteristic of the non-zero
superfluid current. The critical velocity in the STS phase depends
inversely on the values of magnetic field. The critical velocity of
the STS at lower magnetic field is larger than the one in the higher
field. For $J<0$, in SCF phase, the spins in each subsystem are
antiparallel in $xy$ plane. In this phase the translational symmetry
breaks which results in a non-zero transverse staggered
magnetization.  Due to the translational symmetry breaking, four
excitation modes appear in the energy spectrum. Moreover, since the
U(1) symmetry also decreases to $Z_2$ one in this phase the low
energy excitation is gapless with linear dispersion around
$\kv=(0,0)$, and the roton minimum is folded back to the origin.

%%%%%%%%%%%%%%%%%%%%%%%%%%%%%%%%%%%%%%
\section{Anisotropic CAS models}\label{sec:ani-CAS}
%%%%%%%%%%%%%%%%%%%%%%%%%%%%%%%%%%%%%%%

In the previous section we obtained the ground state phase diagram
of the isotropic CAS model in which the inter-chains and
intra-chains NN interactions ($V_1$) and hopping energies ($J$) were
the same. In this section we consider two anisotropic CAS models:
($i$) a CAS system with different inter- and intra-chains NN
interactions, and ($ii$) a CAS system with different inter- and
intra-chains hopping energies, and investigate the effects of these
anisotropies on the stability of the STS order appeared in the
ground state phase diagram of the system. Following, we study these
two anisotropic systems, separately.
%%%%%%%%%%%%%%%%%%%%%%%%%%%%%%%%%%%%%%%%%%%%%%%%%%%%%%%%%%
\begin{figure}[h]
    \centering
    \includegraphics[width=120mm]{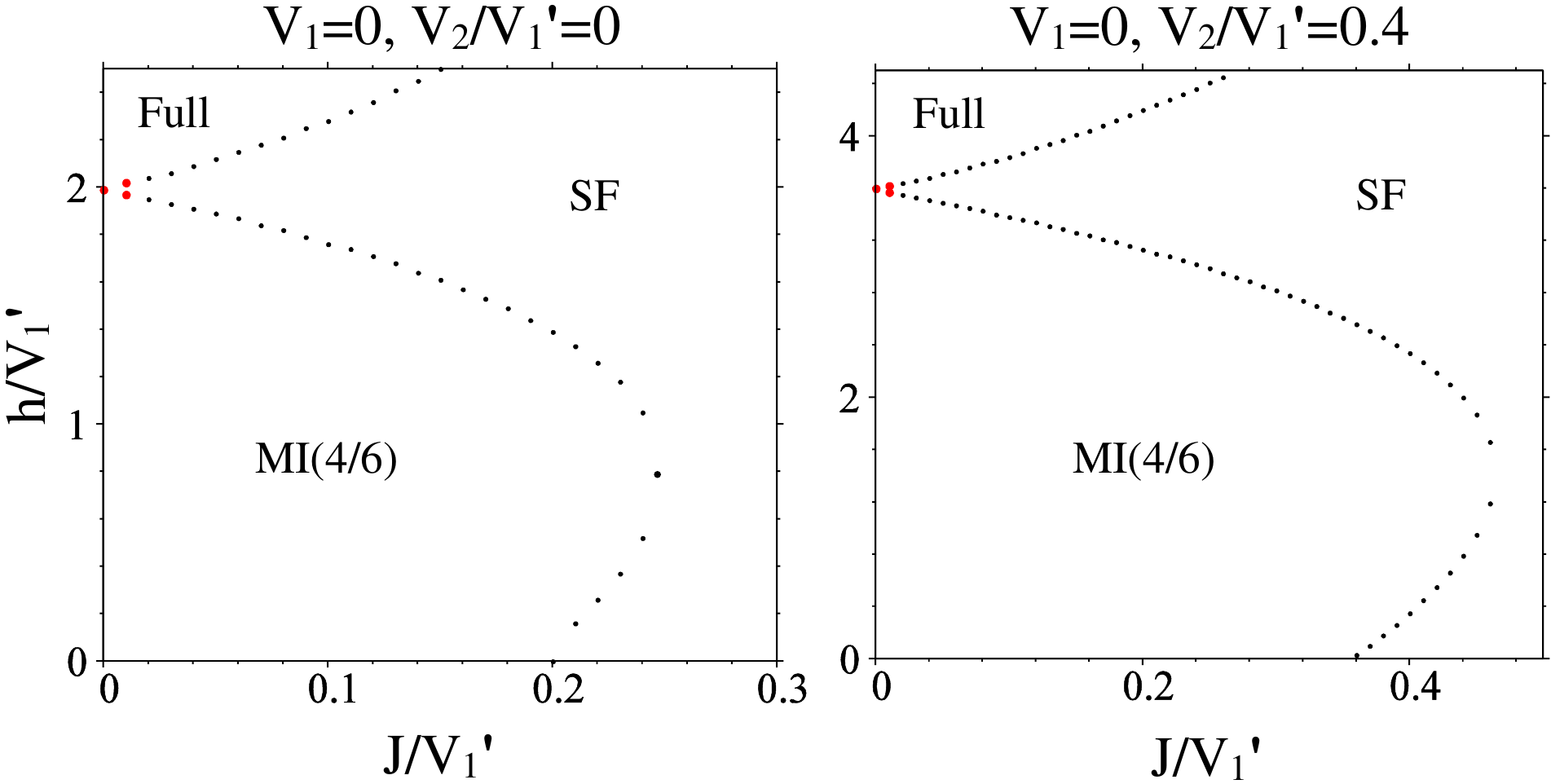}
    \caption{(Color online) CMFT $J-h$ ground state phase diagrams of the anisotropic CAS model in the absence of the intra-chains interactions, ($V_1=0$).
        Independent of the values of NNN interactions, no supersolid order appears in the phase diagram.} \label{fig:diffNN-hj}
\end{figure}
%%%%%%%%%%%%%%%%%%%%%%%%%%%%%%%%%%%%%%%%%%%%%%%%%%%%%%%%%%%%%%%%%%%

\subsection{Anisotropic CAS with different NN interactions}
Let us consider the intra-chains and inter-chains NN interactions to
be respectively $V_1$ and $V_1'$.

\begin{figure}[h]
    \centering
    \includegraphics[width=140mm]{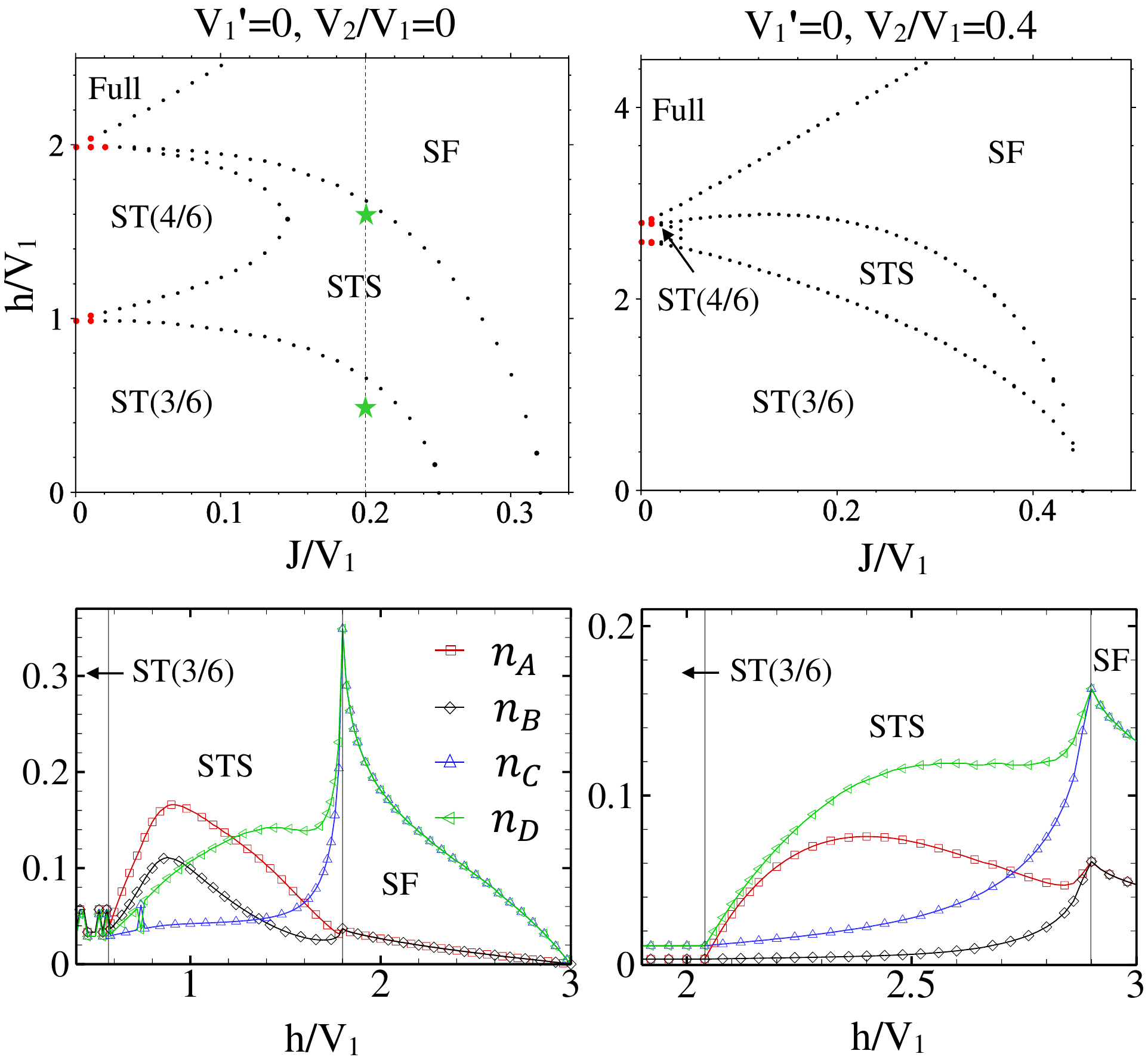}
    \caption{(Color online) Top: CMFT $J-h$ ground state phase diagrams of the anisotropic CAS model in the absence of the inter-chains interactions, ($V_1'=0$). Top-left: in the absence of NNN interactions, ($V_2=0$). Top-right: NNN interactions are $V_2/V_1=0.4$.
        Bottom: Quantum fluctuations (HP bosons' number) on MF ground state, at the line $J/V_1=0.2$, without (bottom-left) and with (bottom-right) NNN interaction.
        Except for near $J\simeq0$, the rest of phase transitions are of second order. Green stars are the SF-STS and STS-ST(3/6) critical points, computed by CMFT-($2\times 4$).} \label{fig:diffNN-hj-STS}
\end{figure}

In the absence of intra-chains interaction, at $V_1=0$, each 1D
lattices are described by a XY Hamiltonian and no supersolid order
appears in the ground state phase diagram of the system. In this
case, when the inter-chains interactions are attractive ($V_1=0,
V_1^\prime <0$), at $h=0$ the system is in the ST(3/6) phase and a
phase transition occurs to the SF phase at $J/|V_1^\prime| \approx
0.23$. In the presence of magnetic field, the system enters the Full
phase (not shown). When the inter-chains interactions are repulsive
($V_1 = 0, V_1^\prime
>0$) the system displays MI(4/6), in addition to the
SF order, but no supersolidity occurs in the phase diagram of the
model (see Fig. \ref{fig:diffNN-hj}). This is due the fact that in
the absence of intra-chains interactions, the translational symmetry
of the chains preserves and the system is always MI(4/6) or SF,
below the saturation field. It should be noticed that the MI(4/6)
phase, appeared in the ground state phase diagram of the isotropic
CAS model (see Fig. \ref{fig:MFphase-diag-U0}), is a result of
competition between NNN interaction and magnetic field. However, in
the mentioned anisotropic CAS model this phase emerges at $h=V_2=0$
where the intra-chains interactions are absent (see top left panel
of Fig. \ref{fig:diffNN-hj-STS}).

%%%%%%%%%%%%%%%%%%%%%%%%%%%%%%%%%%%%%%%%%%%%%%%%%%%%%%%%%%
%%%%%%%%%%%%%%%%%%%%%%%%%%%%%%%%%%%%%%%%%%%%%%%%%%%%%%%%%%%%%%%%%%%
In the absence of inter-chains NN interactions, at $V_1^\prime = 0$,
the anisotropic CAS system is composed of coupled 1D spin-1 and
spin-1/2 XXZ models in a longitudinal magnetic field. In this
system, when the intra-chains interactions are attractive ($V_1'=0,
V_1 <0$), at $h=0$ the system is in the MI(4/6) phase and there is a
phase transition to the SF order at $J/|V_1| \approx 0.26$, and in
the presence of magnetic field, at $h\neq 0$, trivially the system
is in the fully polarized phase (not shown). However, when the
intra-chains interactions are repulsive ($V_1^\prime = 0, V_1
>0$), due to the breaking of the translational and U(1) symmetries, independent of $V_2$, the system exhibits the STS phase. In addition to the STS phase, the ST(3/6) and
ST(4/6) solids, and also SF orders also appear in the phase diagram
of the system (see upper panels of Fig. \ref{fig:diffNN-hj-STS}).
Comparison of the phase diagrams of $V_2=0$ and $V_2\neq 0$ in Fig.
\ref{fig:diffNN-hj-STS} shows that the presence of the NNN
interactions $V_2$ decreases the STS and ST(4/6) regions. It is
surprising that in the absence of $V_2$ the STS phase emerges in the
phase diagram even at zero magnetic field. The amount of quantum
fluctuations in this anisotropic CAS system is plotted in Fig.
\ref{fig:diffNN-hj-STS} (lower panels). As it is seen the
fluctuations in STS phase are not strong and we expect the STS
phases appeared in the CMFT phase diagrams to be stable in the
presence of quantum fluctuations. This achievement is also confirmed
by our CMFT-($2\times 4$) results. According to our CMFT-($2\times
4$) data (not shown), for $V_1'=V_2=0$ although the SF-STS and
STS-ST(3/6) critical points shift to the lower values of magnetic
fields, but the STS region does not become narrower (see green stars
in Fig. \ref{fig:diffNN-hj-STS}).

%%%%%%%%%%%%%%%%%%%%%%%%%%%%%%%%%%%%%%%%%%%%%%%%%%%%%%%%%%
\begin{figure}[h]
    \centering
    \includegraphics[width=140mm]{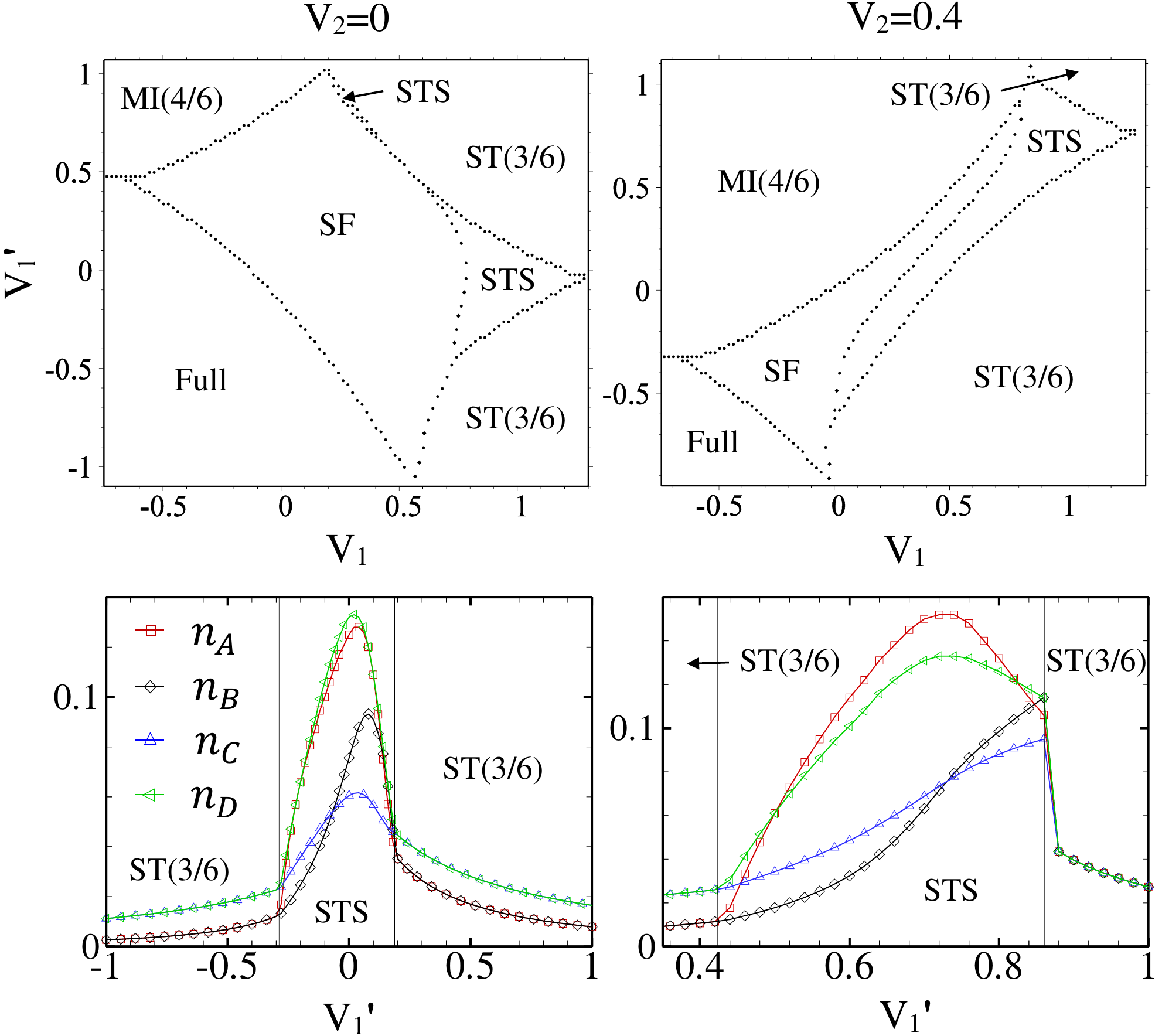}
    \caption{(Color online) Top: CMFT $V_1-V_1'$ ground state phase diagrams of the anisotropic CAS model for $h=1$ and $J=0.2$.
        Bottom: amount of quantum fluctuations at line $V_1=0.9$ (left), and $V_1=0.8$ (right). All transitions are of second order.} \label{fig:diffNN-V1}
\end{figure}
%%%%%%%%%%%%%%%%%%%%%%%%%%%%%%%%%%%%%%%%%%%%%%%%%%%%%%%%%%%%
Above results indicate that, the presence of finite repulsive
intra-chains interactions are necessary for the STS and ST phases to
be emerged in the anisotropic CAS phase diagram. To check this idea
we have also plotted in Fig. \ref{fig:diffNN-V1} the $V_1-V_1'$
phase diagram of the anisotropic CAS system, for the two different
values of NNN interactions: $V_2=0$, and 0.4 at $h=1$ and $J=0.2$.
These figures show that in the presence of attractive intra-chains
interactions, the two Mott insulating phases, MI(4/6) and Full, also
appear in the phase diagram in addition to the SF order. However,
there is no signature of STS and ST phases in this region in both
values of NNN interaction. The STS and ST phases emerge in the
$V_1>0$ region, independent of the strengths of $V_1'$ and $V_2$,
where the translational symmetry breaks in the presence of repulsive
intra-chains interactions. The small amount of quantum fluctuations
in these regions is the reason of the stability of all orders.

From
Fig. \ref{fig:diffNN-V1} it is also seen that in the isotropic CAS
model ($V_1=V_1^\prime$) it is impossible to find the STS phase in
the absence of NNN interactions. Previous studies on the 2D
Bose-Hubbard model with three-body-constrained bosons, show solid
and superfluid phases in the presence of isotropic NN interactions
and hopping terms \cite{chen2011quantum}, but no supersolidity
occurs in this system. Actually, NNN interactions can stabilize the
supersolid phases in this model.

\subsection{Anisotropic CAS with different hoppings}
\begin{figure}[ht!]
    \centering
    \includegraphics[width=75mm]{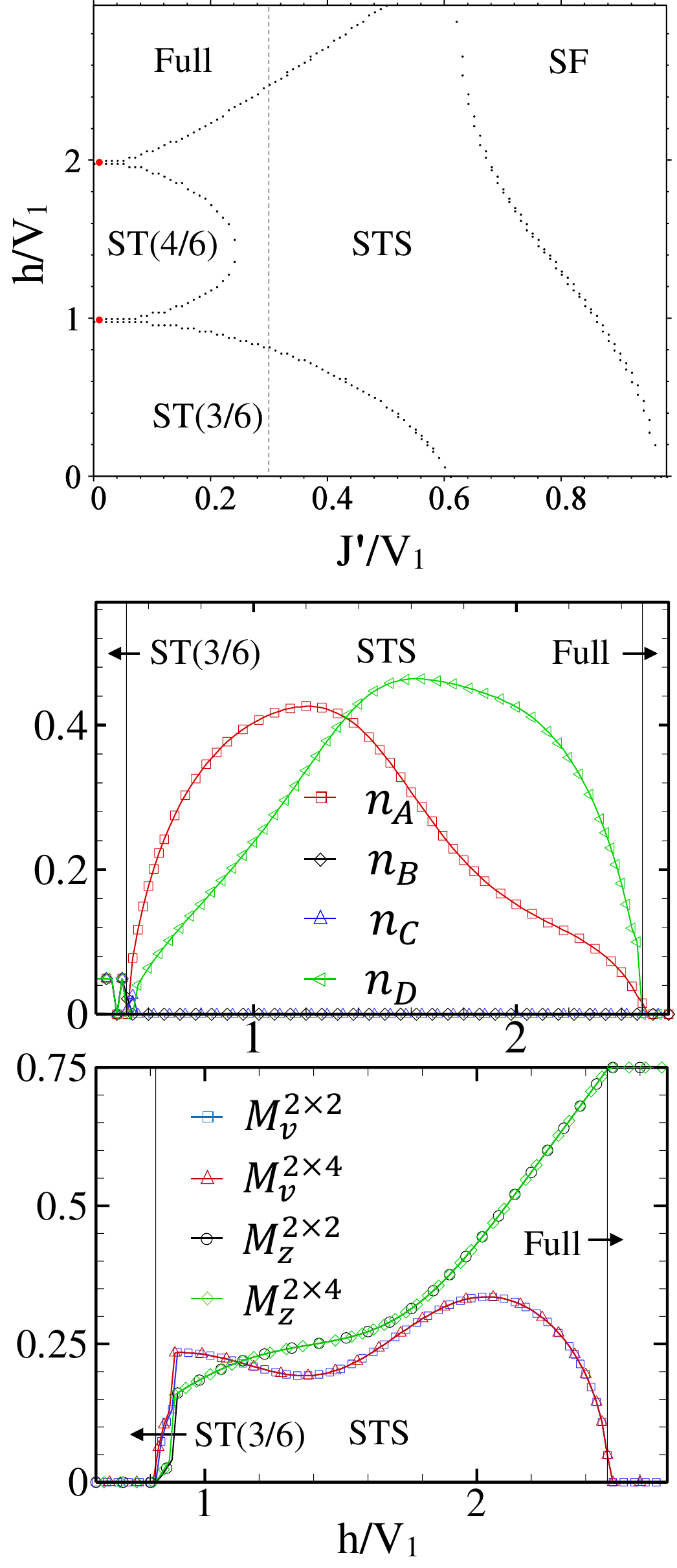}
    \caption{(Color online) Top: CMFT $J'-h$ ground state phase diagram of the anisotropic CAS model ($V_1'=V_2=0$). Middle: amount of quantum fluctuations at line $J'/V_1=0.3$, shown in the phase diagram.
        Bottom: Order parameters computed by CMFT-($2\times 2$) and CMFT-($2\times 4$) at the same line.} \label{fig:diff-j}
\end{figure}

In this subsection we consider an anisotropic CAS model with
different inter- and intra-chains hopping energies and investigate
the effects of this anisotropy on the ground state phase diagram of
the system. Let us consider the intra- and inter-chains hopping
energies to be respectively $J$ and $J'$. Suppose one of the hopping
energies $J$ or $J'$ to be zero. Our CMFT results show that the
presence of inter-chains hoppings together with the repulsive
intra-chains NN interactions are sufficient for the appearance of
STS phase (see Fig. \ref{fig:diff-j}). The behavior of HP bosons'
number shows that in the STS phase the quantum fluctuations are
exactly zero in the sublattices B and C, but considerable in A and
D. This means that we should expect some modifications on the
sublattices' magnetizations in the presence of fluctuations.
However, surprisingly we see that the order parameters computed by
CMFT-($2\times 2$) and -($2\times 4$) are exactly the same (see Fig.
\ref{fig:diff-j}) which is an indication of the stability of the STS
phase in this system.

%%%%%%%%%%%%%%%%%%%%%%%%%%%%%%%%%%%%%%%%%%%%%%%%%%%%%%%%%%
%%%%%%%%%%%%%%%%%%%%%%%%%%%%%%%%%%%%%%%%%%%%%%%%%%%%%%%%%%%%%%%%%%%
\subsection{Minimal model}
Based on the above results we conclude that, in the presence of
intra-chains hoppings, the minimal mixed-spin CAS model for the
supersolidity is given by the following Hamiltonian:
\begin{equation}
H=H^{\s}_{XXZ}+H^{\t}_{XXZ}+H^{\s\t}_{XY}, \label{minimal-Hamiltonian}
\end{equation}
where, $H^{\s}_{XXZ}$ and $H^{\t}_{XXZ}$ are respectively the
spin-1/2 and spin-1 XXZ Hamiltonians, and $H^{\s\t}_{XY}$ is the XY
Hamiltonian, which couples the spin-1/2 and spin-1 chains. In
bosonic language, the presence of off-site intra-components
interactions and hopping energies together with the inter-components
hoppings are sufficient to find the STS phase in the two-component
hard-core Bose-Hubbard model. Moreover, in the absence of NN
intra-chains hoppings, the minimal mixed-spin CAS model for the
supersolidity is, instead, spin-1 and spin-1/2 Ising chains
interacting via a XY Hamiltonian.

%%%%%%%%%%%%%%%%%%%%%%%% Summary %%%%%%%%%%%%%%%%%%%%%%%%
\section{Summary and conclusion}\label{sec:summary}

To summarize, in the present paper, employing three analytical and
numerical approaches, MF approximation, CMFT with different cluster
sizes and  LSWT, we have studied the ground state phases of a 2D
mixed-spin system of coupled alternating spin chains described by
the spin Hamiltonian in Eq. (\ref{Hamiltonian}). Our study,
indicates that the CAS system displays a rich ground state phase
diagram including STS and SCF phases in addition to the different
solids, SF and MI phases.  We have also considered two kinds of
anisotropic CAS model, ($i$) CAS model with different intra-chains
and inter-chains NN interactions and ($ii$) CAS model with different
intra-chains and inter-chains hoppings, and investigated the effects
of these anisotropies on the ground state phases. We have
demonstrated that the emergence of the STS phase strongly depends on
the strength of intra-chains NN interactions and hopping energies.
We have shown that, for the systems with uniform hoppings, the
repulsive intra-chains interactions are necessary and sufficient for
stripe supersolidity. In this case the minimal two dimensional
mixed-spin model is a system of spin-1 and spin-1/2 XXZ chains,
interacting via a XY Hamiltonian. However, in the case of
anisotropic hoppings, the STS phase emerges even in the absence of
intra-chains interactions, and a system of coupled Ising chains is
the minimal model.

Our mixed-spin model is equivalent to a bosonic system of hard-core
and semi-hard-core bosons and could be realized in coupled one
dimensional optical lattices by alternatively changing the optical
depth. Study of temperature phase diagram as well as thermodynamic
properties of the CAS system and also study of the ground state
phase diagram with other approaches are left for future work.

%%%%%%%%%%%%%%%%%%%%%%%%%%%%%%%%%%%%%%%%%%%%
%%%%%%%%%%%%%%%%%%%%%%% References %%%%%%%%%%%%%%%%%%%%%%%%%%%%%%%%%%%%
\newpage
\bibliographystyle{plain}

%

%\section*{Author contributions statement}

%F. H. performed the calculations. J. A. wrote the manuscript. All
%authors approved the final manuscript. J. A. supervised the
%research.

%\section*{Additional information}

%\textbf{Competing financial interests}: The author declares no competing financial interests.\\
\end{document}